\documentclass[preprint,showpacs,superscriptaddress]{revtex4}
\usepackage{graphics,epsfig}
\usepackage{graphicx}
\usepackage{dcolumn}
\usepackage{amsmath}
\usepackage{epstopdf}
\newcommand{\be}{\begin{equation}}
\newcommand{\ee}{\end{equation}}
\newcommand{\beq}{\begin{equation}}
\newcommand{\eeq}{\end{equation}}
\newcommand{\bea}{\begin{eqnarray}}
\newcommand{\eea}{\end{eqnarray}}
\newcommand{\nn}{\nonumber}
\def\be{\begin{equation}}
\def\ee{\end{equation}}
\def\ba{\begin{eqnarray}}
\def\ea{\end{eqnarray}}

\begin{document}
\title{Logarithmic Corrections to the  Entropy of Scalar Field in BTZ Black Hole Space-time}

\author{ Dharm Veer Singh\footnote{Present Address: Centre for Theoretical Physics, Jamia Millia Islamia, New Delhi} and Shobhit Sachan\footnote{Present Address: Government Engineering College, Mainpuri.}}

\address{Physics Department, Centre of Advanced Studies,\\
 Banaras Hindu University, Varanasi - 221 005, (U.P.), India\\
e-mails: $^\star$veerdsingh@gmail.com, $^\dagger$shobhitsachan@gmail.com} 
\begin{abstract}
 The entanglement entropy correlates two quantum sub-systems which are the part of the larger system.  A logarithmic divergence term  present in the  entanglement entropy is universal in nature and directly proportional to the conformal anomaly. 
 We study this  logarithmic divergence term of entropy for massive scalar field in $(2+1)$ dimension by applying numerical techniques to entanglement entropy approach.
 This (2+1) dimensional massive theory can be obtained from (3+1) dimensional massless scalar field via dimensional reduction.  We also calculated mass corrections to entanglement entropy for scalar field. Finally, we observe that the area law contribution to the entanglement entropy is not affected by this mass term and the universal quantities depends upon the basic properties of the system.
\end{abstract}
\maketitle

\section{\label{sec:1}Introduction}
Black holes are gravitational solutions of Einstein's field equations.  Black holes have some properties similar to that of a thermodynamical system.  Therefore, like a  thermodynamical system, entropy and temperature can be assigned to black holes. The temperature of a black hole is directly proportional to surface gravity of the event horizon.The entropy,  known as Bekenstein-Hawking entropy is directly  proportional to the area of the event horizon\cite{JD1,JD2,JD3,JD4,SWH}.

 There are many attempts made  to understand the origin of the  black hole entropy. Some of the examples of these attempts are  based on  the calculation of {\bf a)}   the  value of  Euclidean action \cite{7, 8, 10}, {\bf b)} the rate of the pair creation of black holes \cite{11}, {\bf c)} the Noether charge of the bifurcate Killing horizon \cite{13,14} and {\bf d)} the central charge of the Virasoro algebra \cite{15, 16, 17}. 

The microscopic  derivation of the black hole entropy was given in superstring theory \cite{d1,d2} by using the so-called D-brane method  \cite{dbrane,d3,d4}.  In 1985,  `t Hooft  introduced another  model to calculate the  entropy  of a black hole, known as the brick wall model \cite{thooft}.  Beside of all these  previously well studied models, we concentrate our study on the  entanglement entropy model  \cite{Bombelli,19, 20, 21, 22, 23,24, 25,Cadoni:2007vf,Cadoni:2009tk,Cadoni:2010vf,Hung:2011ta}  as this is the  most attractive candidate for the black hole entropy. 

The entanglement entropy is the source of  quantum information. It is a  measure of the correlation between  subsystems, separated by a boundary called the entangling surface \cite{Bombelli,19}.  It is also a  measure of the information loss due to division of the system. The entanglement entropy depends upon the geometry of the boundary, but not on the properties of the subsystems. The entanglement entropy is defined by the von Neumann entropy.

 We study the logarithmic contribution to the entropy for scalar fields by using the dimensional reduction technique. In this technique, the coefficient of logarithmic divergence term in (2+1) dimensional massive theory can be obtained via dimensional reduction of (3+1) dimensional massless theory ($\Sigma_2=\Sigma_1\times S^1$) using the entanglement entropy method. The reduced density matrix, which arises in the formulation, are written in terms of correlators \cite{Casini:2009sr}.  The reduced density matrix in terms of correlators is well known for scalar fields and obeys the Wick's theorem. The logarithmic divergence terms in  entropy of black holes appear  due to the  infinite number of states near the horizon and these divergences scaled by the size of the black holes. These logarithmic divergence terms are related to the conformal anomalies.  In even dimensions, conformal field theory (CFT) contains a divergence term, but in odd dimensions there is no divergence term across the entangling surface \cite{MPH,Hertzberg:2012mn}.  The coefficient of logarithmic term is proportional to the conformal anomaly \cite{SOLO1} ($a$ and $c$ type anomaly). For a spherical system, the results of  ``$a$'' type anomaly  can be extended in any dimension \cite{Myers1,Hung:2011xb}, but ``$c$'' type anomaly can not be extended in higher dimensional theory \cite{SOLO2}.  

This paper is organized as follows; we have given brief review of free massive theory in Sec. 2. We study the  scalar field   in BTZ black hole space-time in Sec. 3  and numerical calculations for logarithmic contribution to the entanglement
entropy  in Sec. 4. We present our results and their physical implication of entropy for scalar fields in BTZ black hole space-time in Sec 5. Some formulas which are used in the text, are defined in Appendix A.
\section{Free Massive Theory in BTZ Space-Time}
\label{freetheory}
The general structure of entanglement entropy of the system  with logarithmic  divergence is given by the relation,
\be
S=\frac{A}{4\pi\epsilon^2}+s\ln\epsilon,
\ee
 where $s$ is the coefficient of the logarithmic divergence term and $\epsilon$ is the  ultraviolet cutoff. The first part, which is finite, is Bekenstein-Hawking area law and second one is logarithmically divergence term of the entanglement entropy. For general  conformal field theories in $(3+1)$ dimensions, the logarithmically divergence term is directly related to the $a$ and $c$ conformal  anomalies. The relation between logarithmically divergence term and $a$ and $c$ type of anomalies is given\cite{H2}, 
\be
s=\frac{a}{180}\chi(\partial V)+\frac{c}{240\pi}\int_{\Sigma_2}(k_i^{\mu\nu}k^{i}_{\nu\mu}-\frac{1}{2}k_i^{\mu\mu}k^{i}_{\mu\mu}),
\label{general}
\ee
where $\chi(\partial V)$ is the Euler number of the surface $\Sigma_2(\Sigma_2=\Sigma_1\times R$ where $R$ is radius of the cylinder).  The $k^i_{\mu\nu}=-\gamma^\alpha_\mu \gamma^\beta_\nu \partial_\alpha n^i_\beta$ is the extrinsic curvature, $n^\mu_i$ (with $i=1,2$) are the pair of unit vector orthogonal to $\partial V$, and $\gamma_{\mu\nu}=\delta_{\mu\nu}-n^i_\mu n^i_\nu$ is the induced metric on the surface. From  equation (\ref{general}), we can see that the coefficient of the logarithmic divergence of the cylinder is proportional to the $c$ type anomaly and is given by,
\be
s=\frac{c}{240}\,\frac{L}{R},
\label{general1}
\ee 
where $R$ and  $L$ are the radius  and length of the   of the cylinder.
Let us consider a system of three spatial dimensions $x_1,x_2, \mathrm{and}~ x_3$ of the form $\Sigma=\Sigma_2\times x_1$. The direction $x_1$ can be compactified by imposing the boundary conditions and thus the system reduces to two dimensions. The Fourier decomposition of the corresponding field modes in the compactified direction is given by,
\be
\Phi\,(t,r,\theta,\phi)=\phi\,(t,r,\theta)\exp\left[i\frac {2\pi m}{L}\phi\right].
\ee
This decomposition of fields enable us to write the EOMs in form,
\be
\partial_\mu^2\phi+M_m^2\phi=0
\ee 
where
\be
M^2_m=\mu^2+(\frac{2\pi}{L}m)^2.
\label{A1}
\ee
In above definition,  $\mu$ is the mass of the free fields and acts as infrared correlator and $m$ is an azimuthal quantum number. In our study, we consider  the free massless field, therefore  we set  $\mu=0$. In this case, the equation (\ref{A1}) becomes \cite{Safdi1,Safdi2},
\be
M^2_m=(\frac{2\pi}{L}m)^2,
\ee
The contribution of entanglement entropy of the two dimensional fields is given by the relation \cite{H2},
\ba
S( \Sigma)&=&\sum_{m=-\infty}^{\infty} S\,(\Sigma_2 ,M_m)  =\frac{L}{\pi}\int_0^{\infty}dM\, S(\Sigma_2,M).
\label{general2}
\ea
We expand $S(\Sigma_2,M)$ in terms of $(MR)$ and neglect higher order terms, obtaining the relation,
\be
S(\Sigma_2,M)=c_0+c_1\,MR+\sum_{n=0}^{\infty}\frac{c_{-1}}{MR}
\ee
Substituting the value of $S(\Sigma_2,M)$  in equation (\ref{general2}), we obtain the  the logarithmic coefficient $s$  in $S(\Sigma )$ which is directly related to $c_{-1}$ by  the relation,
\be
s_s=c_{-1}\frac{L}{\pi R}
\ee
The coefficient $c_{-1}$ is  obtained from the free massless theory in (3+1) dimensions and is directly related to the coefficient of $(MR)^{-1}$ . The coefficient $c_{-1}$ is found $-\frac{\pi}{240}$ for scalars. 

Now, The logarithmic divergence term of entropy is proportional to the mass term in the dimensionally reduced theory and given by the term $c_1$. The entropy of scalar field is given by \cite{MPH},
\be
\Delta S_{M}=\gamma_d M^{d-1}A_{d-1},
\ee
where
$\gamma_d\equiv(-1)^{(d/2)}[12 ~(2\pi)^{(d-2)/2}(d-1)!!]^{-1}
$
and ``$A$'' is the area of event horizon $(A=2\pi r_+)$. For $d=2$, the value of $\gamma_d$ is  $\gamma_2=-\frac{1}{12}$. The coefficient $c_1$ is linear with entropy and it is found $-\frac{2\pi}{12}$.

\section{\label{see:level2} Scalar Fields in BTZ Black Hole Space-time}
Let us consider the action of the (2+1) dimensional gravity with cosmological constant $\Lambda$ \cite{MB,Carlip},
\be
S=\frac{1}{2\pi}\int d^3x\sqrt{-g}\,[R+2\Lambda].
\ee
The value of cosmological constant is ${-\frac{1}{l^2}}$. One of the solution of this $(2+1)$ dimensional gravity with negative cosmological constant is is BTZ black hole. The metric of BTZ black hole is given by the equation;
\be
{ds}^2=-(-M+{r}^2/{l}^2){dt}^2+\frac{1}{(-M+{r}^2/{l}^2+{J}^2/{4r}^2)}{dr}^2+{r}^2{d\phi}^2-J{dt}{d\phi}~,
\ee
where $-\infty < t < \infty$ and $0\leq\phi\leq2\pi$. The metric of the BTZ black hole in term of proper distance $r^2=r_+^2\cosh ^2\rho+r_-^2\sinh ^2\rho$ is written as, 
\ba
ds^2=-\Big(u^2+\frac{J^2}{4l^2(u^2+M)}\,\Big)\,dt^2+d\rho^2+(\frac{J}{2l\sqrt{(u^2+M)}}dt-l\sqrt{{u^2+M}}d\phi)^2\nn\\.
\label{pd}
\ea
Where $r(\rho)^2=l^2(u^2+M)$ and $r_+$ and $r_-$ are inner and outer horizon of the black hole respectively.

The action of massive scalar field in the background of BTZ black hole is written as,
\be\label{eq:action1}
S=-\frac{1}{2}\int dt\sqrt{-g}\,(g^{\mu\nu}\,\partial_{\mu}\Phi\partial_{\nu}\Phi+\mu^2\Phi^2)
\ee 
where $\sqrt{-g}$ and $g^{\mu\nu}$ are the determinant and the metric element of the BTZ black hole (\ref{pd}).  The field $\Phi$ can be decomposed using the separation of variables $\Phi({t,\rho,\phi})=\sum_m\,\phi_m(t,\rho)\,e^{im\phi}$. This decomposition of $\Phi$ manifest the cylindrical symmetry of the system. Substituting the value of $\Phi,             g^{\mu\nu}$ and $\sqrt{-g}$ in equation \eqref{eq:action1}, we get the expression,
\ba
S&=&-\frac{1}{2}\int dt\Big[-\frac{\sqrt{(u^2+M)}}{\sqrt{[u^2+\frac{J^2}{4(u^2+M)}]}}\,\dot{\Phi_m}^2+u\sqrt{[(u^2+M)+\frac{J^2}{4u^2}]}(\partial_\rho\Phi_m)^2\nn\\&+&\frac{u^2m^2}{u\sqrt{[(u^2+M)+\frac{J^2}{4u^2}]}}{\Phi_m}^2-\frac{(iJm)}{u\sqrt{[(u^2+M)+\frac{J^2}{4u^2}]}}\Phi_m\dot{\Phi_m}\nn\\&&+\mu^2{\sqrt{[u^4+u^2M+\frac{J^2}{4}]}}\Phi^2\Big]\nn\\,
\ea
where $\pi_m$ is the conjugate momentum corresponding the field $\phi_m$. The Hamiltonian of the scalar field in the BTZ background space-time is  given by \cite{DS1,DS2,DS3},
\begin{align}
&H=\frac{1}{2}\int d\rho\,\left({\frac{{u^2+\frac{J^2}{4\left(u^2+M\right)}}}{{\left(M+u^2 \right)}}}\right)^{1/2}\pi^2+\frac{1}{2}\int\,d\rho\, d\rho'\,u\sqrt{\left[(u^2+M)+\frac{J^2}{4u^2}\right]}\,\partial_{\rho}(\Phi_m)^2\nn\\
&+\left({\frac{{u^2+\frac{J^2}{4(u^2+M)}}}{{(M+u^2 )}}}\right)^{1/2}\,\Phi_m(\rho)^2-iJ\left({\frac{{1}}{{(M+u^2 )}}}\right)^{1/2}\,\Phi_m(\rho)\,{\dot \Phi}_m(\rho)\nn\\&+\mu^2{\sqrt{[u^4+u^2M+\frac{J^2}{4}]}}\Phi^2
\label{hamilt1}
\end{align}
This Hamiltonian is not diagonal, therefore to diagonalize it,  we define the new momentum $\tilde{\pi}_m=\pi_m-{iJm}/{u({[(u^2+M)+{J^2}/{4u^2}]}})^{1/2}\Phi_m $. The canonical variables, the field ($\phi_m$) and diagonalized momentum ($ {\tilde \pi}_m$) satisfy the following relation,
\be
[\phi_m(\rho), {\tilde \pi}_m(\rho)]=\frac{i\,J\,m}{u\sqrt{\left[(u^2+M)+\frac{J^2}{4u^2}\right]}}\delta_{m,m^{\prime}}\delta(\rho-\rho^{\prime}).
\label{com}
\ee
Using the diagonalized momentum, the diagonalized Hamiltonian can be written as \cite{DS1,DS2,DS3},
\begin{align}
H=&\frac{1}{2}\int d\rho\,\tilde{\pi}_{m}^2(\rho)+\frac{1}{2}\int\,d\rho\, d\rho'\,u\sqrt{\left[\left(u^2+M\right)+\frac{J^2}{4u^2}\right]}\,\nn\\
&~~~~~~~~~ \left(\partial_{\rho}\left(\sqrt{\frac{\sqrt{\left[u^2+\frac{J^2}{4(u^2+M)}\right]}}{\sqrt{(u^2+M)}}}\right)\,\psi_m\right)^2+\mu^2{\frac{{u^2+\frac{J^2}{4(u^2+M)}}}{{(M+u^2 )}}}\psi_m^2,
\label{hamilt2}
\end{align}
where
\be
\psi_m(t,\rho)=\left({\frac{{u^2+\frac{J^2}{4(u^2+M)}}}{{(M+u^2 )}}}\right)^{1/4}\,\Phi_m(t,\rho),
\ee
For general quadratic case, the Hamiltonian of the system is written as,
\be
H=\frac{1}{2}\sum\pi^2+\frac{1}{2}\sum_{AB}V_{AB}q^Aq^B
\ee
where $q_A$ and $\pi_A$ obey the commutation relation $[q^A,\pi^B]=i\,\delta^{AB}$ and  $V_{AB}$ is the matrix. The two point correlator is given by,
\ba
&&X_{ij}=\langle q_A\,q_B\rangle=\frac{1}{2}(q_A\,q_B)^{1/2}=(M+N)^{-1}_{AB}\nn\\
&&P_{ij}=\langle \pi_A\,\pi_B\rangle=\frac{1}{2}(\pi_A\,\pi_B)^{1/2}=(M-N)_{AB}
\ea
where $M$ and $N$ are defined in appendix (A). Then the entropy of the system is given by the relation
\be
S\,\Big(R=(n_{B}+\frac{1}{2})a\Big)=\lim_{n\rightarrow\infty}S(n,N)=S_0+\sum_m\,S_{ent}^m. 
\ee
where
\begin{align}
S_{ent}&=\mathrm{Tr}\left(\left(\sqrt{X_mP_m}+\frac{1}{2}\right).~\log\left(\sqrt{X_mP_m}+\frac{1}{2}\right)\right.\nn\\
&~~~~\left.-\left(\sqrt{X_mP_m}-\frac{1}{2}\right).~\log\left(\sqrt{X_mP_m}-\frac{1}{2}\right)\right).
\end{align}
 We make the following replacements to discretized the Hamiltonian of the system,
\ba
u[\rho=(i-\frac{1}{2})\alpha]\rightarrow u_i, \qquad\qquad\Psi_m[\rho=(i-\frac{1}{2})\alpha]\rightarrow\Psi^{i}_m, \nn
\ea
where $i,~j=1,2 ....N$ and ``$\alpha$'' is the lattice spacing . We discretized Hamiltonian (19) using the above replacements and we suppressed the angular momentum index $m$. The matrix elements $V_{AB}$ corresponding to the discretized Hamiltonian (19) is given by, \footnote{We discretized the Hamiltonian (\ref{hamilt2}) using the the middle-point prescription and the derivative of the form $f(x)\partial g(x)$ is replaced by $\frac{f_{j+1/2}[g_{j+1}−g_j]}{\alpha}$},
 \begin{align}
V_{AB}^m \psi_m^A\psi_m^B=&a\sum_{A=1}^N\left[u_{A+\frac{1}{2}}\sqrt{(u^2_{A+\frac{1}{2}}+M)}\left(\frac{\sqrt{u^2_{A+1}}}{\sqrt{u^2_{A+1}+M}}\psi_m^{A+1}\-\frac{\sqrt{u^2_A}}{\sqrt{u^2_A+m}}\psi_m^{A}\right)^2\right.\nn\\
&~~~~~~~~~~~~~~+\left.\frac{m^2}{r_+}\,\frac{u^2_A}{u^2_A+M}\psi_m^{A^2}+\mu^2\Phi_m^2\right].
\label{mat}
\end{align}
The diagonal and off-diagonal terms are given by,
\begin{align}
\Sigma_A^{(m)}=&\frac{\sqrt{\left[u^2_{A}+\frac{J^2}{4(u^2_{A}+M)}\right]}}{\sqrt{\left(u^2_{A}+M\right)}}\left(u_{A+1/2}\sqrt{(u^2_{A+1/2}+M)+\frac{J^2}{4u^2_{A+1/2}}}\right.\nn\\
&-\left.u_{A-1/2}\sqrt{(u^2_{A-1/2}+M)+\frac{J^2}{4u^2_{A-1/2}}}~~\right)+m^2{\frac{{u^2_A+\frac{J^2}{4(u^2_A+M)}}}{{(M+u^2_A )}}},\\
\Delta_A=&-u_{A+1/2}\sqrt{\left(u^2_{A+1/2}+M\right)+\frac{J^2}{4u^2_{A+1/2}}}\nn\\
&~~~~~~~~~~\sqrt{\frac{\sqrt{\left[u^2_{A+1}+\frac{J^2}{4(u^2_{A+1}+M)}\right]}}{\sqrt{(u^2_{A+1}+M)}}}\sqrt{\frac{\sqrt{\left[u^2_{A}+\frac{J^2}{4(u^2_{A}+M)}\right]}}{\sqrt{\left(u^2_{A}+M\right)}}}
\end{align}
where $A,~B=1,2 ....N$ and ``$a$'' is UV cut-off length. We regain the continuum by taking the limit $a\rightarrow 0$ and $N\rightarrow \infty$ while the size of the system remains fixed.
\section{\label{sec:level3}Numerical Estimation}
In this section, we study the  numerical estimation of entanglement entropy of massive scalar field in BTZ black hole space time. We start from with the calculation of $(N\times N)$ matrix of $ V_{AB}$, where $AB=1,2\ldots N$ for given  mass $(\mu)$ and angular momentum $(m)$. We calculate the correlator $X_{ij}$ and $P_{ij}$ and then calculate the entropy of massive field in BTZ black hole space-time. For the numerical computation, we consider the system is discretized in radial direction with  lattice size $N=200$  and the partition size $n_B=10,20\ldots 100$.

 The entropy can be expanded in powers of proper distance, $\rho$, for large values of $\rho$,
\be
S=c_0(M)+c_1(M)\,\rho+c_{-1}(M)\frac{1}{\rho}+\ldots.
\ee

\begin{figure}[t]
\centering
\includegraphics[width=0.8\textwidth]{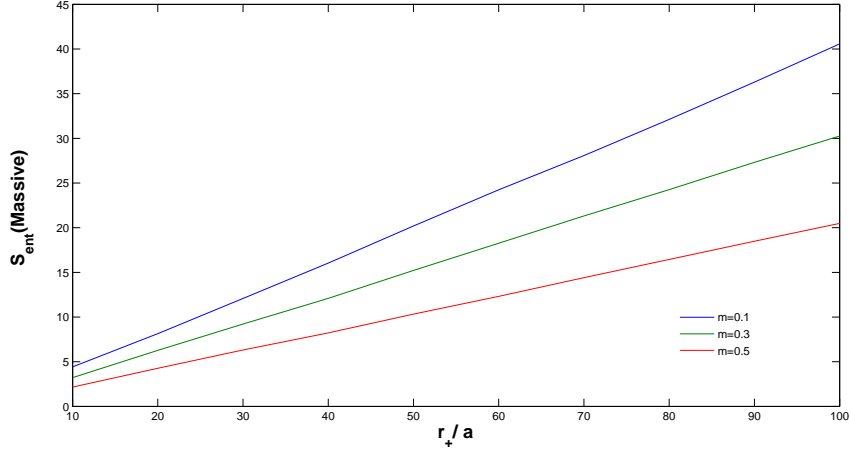}
\caption{The numerical calculation for $S_{ent}(massive)$ of the scalar field in rotating BTZ space-time. $S_{ent}$ is shown as a functions of $r_+/a$ for different masses m=0.1, 0.3 and 0.5 . We have taken the lattice point N=200.}
\label{fig:mass}
\end{figure}

The entropy of  scalar field for different masses in the range $(.05<M<.5)$ is computed numerically. The Value of $c_1(\mu)$ and $c_{-1}(\mu)$ are tabulated in the table (\ref{tab:mass}),
\begin{center}
\begin{table}[h]
\begin{center}
\begin{tabular}{|l|l|r|l|r|l|r|}
\hline
\multicolumn{1}{|c|}{ } & \multicolumn{1}{c|}{$M=0.1$ } & \multicolumn{1}{c|}{ $M=0.2$ }& \multicolumn{1}{c|}{$M=0.3$} & \multicolumn{1}{c|}{$M=0.4$}& \multicolumn{1}{c|}{$M=0.5$}\\
\hline

\,\, $c_1(M)$\,\, & \,\, 0.401\,\, &0.354\,\, & \,\, 0.302& \,\, 0.241& \,\, 0.204
\\
\,\, $c_{-1}(M)$\,\, & \,\, 0.200\,\, &0.068\,\, & \,\, 0.050& \,\, 0.040& \,\, 0.030
\\
\hline
\end{tabular}
\end{center}
\caption{The value of $c_0,c_1$ and $c_{-1}$ for different masses 0.1, 0.2, 0.3, 0.4 and 0.5.}
\label{tab:mass}
\end{table}
\end{center}
The value of coefficients $c_1(M)$ and $c_{-1}(M)$ are tabulated in the table (\ref{tab:mass}) and shown in figure (\ref{fig:mass}). If we calculated the value of coefficients, then we expand the $c_1(M)$ and $c_{-1}(M)$ in power of M,
\ba
&&c_1(M)=c_1^1(M)+c_1^0+c_{1}^{-1}\frac{1}{M},\\
&&c_{-1}(M)=c_{-1}^1(M)+c_{-1}^0+c_{-1}\frac{1}{M}
\ea

\begin{figure}[htp]
\centering
\includegraphics[width=0.8\textwidth]{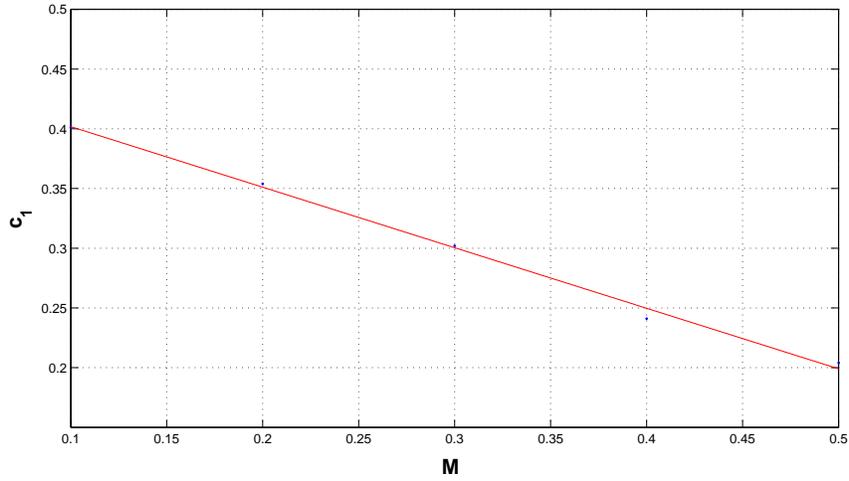}
\caption{The points corresponds to the coefficient of the linear term
in r in the entanglement entropy for different masses. }
\label{fig:2}
\end{figure}

\begin{figure}[h]
\centering
\includegraphics[width=0.8\textwidth]{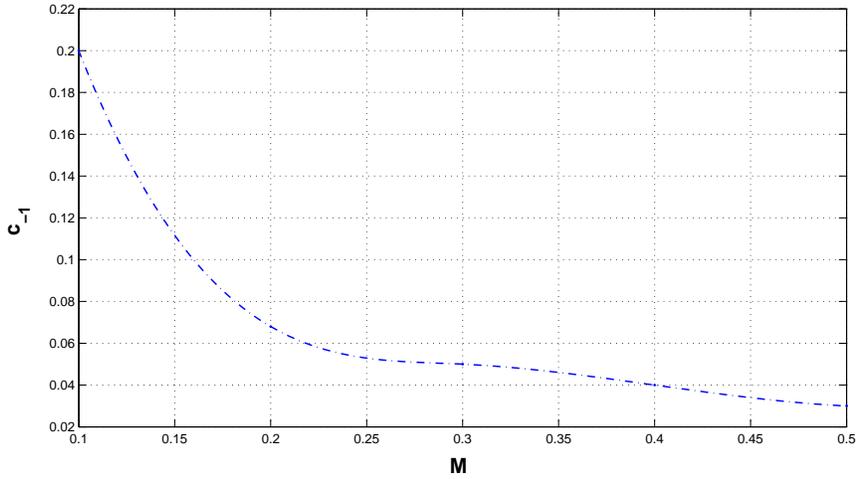}
\caption{The points corresponds to the coefficient of the term $1/{\rho}$ in
the entanglement entropy for different masses.The coefficient of term is proportional to
1/M in the fit drawn with a solid line is 0.0132 = $\pi$/240. This is the value of $c_1$ in
equation (5.22).}
\label{fig:3}
\end{figure}

The plot $c_1$ and  $c_{-1}$  as the coefficient  of $M$ and $1/M$ in (29) and (30). The value of $c_1(M)$ and  $c_{-1} (M)$ as shown in figure (\ref{fig:2}) and  (\ref{fig:3}). The co-efficients $c_1$ and $c_{-1}$ are found from the fitting the data
plotted in figure (3) and the values are -0.503 and -0.0132 respectively. Here it is  interesting to note that the  co-efficients  $c_{-1}$ is related with the co-efficient of logarithmic term in (3+1) dimension and is given by $-\pi/240$ and for the  co-efficients  $c_{1}$ (is obtained from dimensionally reduced theory ) is given by $-\pi/6$.


\section{\label{sec:level4}Conclusion}
In this paper, we have studied the logarithmic divergence term of entanglement entropy for the scalar field propagating in the background of BTZ black hole numerically. The coefficient of divergence term $c_1$ and $c_{-1}$ calculated numerically. The logarithmic divergence term(s) of entanglement entropy is the linear combination of $c$ type anomaly. The term $c_1$ is obtained from the dimensional reduction of the  theory and the term $c_{-1}$ is directly related to the coefficient of divergence term. The general structure of the coefficients is same as that found in (\ref{general2}). This is the agreement of our numerical results with analytical results \cite{MPH}.  We can also extend our results for the higher dimension theory. We have
also studied the logarithmic divergence term of entanglement entropy for the fermion field propagating in the background of BTZ black hole numerically \cite{DS4}.

\appendix
\section{Model of entanglement Entropy}
In this appendix, the model of entanglement entropy for scalar field and numerical computation of entropy is reviewed. Let there is a system of coupled harmonic oscillators $q^A,~ (A=1,.......,N)$ which one can use to study the entanglement entropy of the system. The Hamiltonian of this coupled harmonic oscillator system is written as,
\be
H=\frac{1}{2a}\delta^{AB}p_Ap_B+\frac{1}{2}V_{AB}q^Aq^B,
\ee
Where $p^A$ and $p^B$ are canonical momentum  corresponding to the $q^A$ and   $q^B$ respectively. Tha canonical momenta are given by the relation $p_A=a\,\delta_{AB}\,\dot{q}^B$, where $\delta_{AB}$ is Kronecker delta,  $V_{AB}$ is real, symmetric, positive definite matrix and ``a''  is fundamental length characterizing the system. Using the creation and annihilation operators, one can write total Hamiltonian as,
\be
H=\frac{1}{2a}\delta^{AB}\left(p_A+iW_{AC}q^C\right)\left(p_B-iW_{BD}q^D\right)+\frac{1}{2a}\,\mathrm{Tr}~ W
\ee
where $W$ is symmetric, positive definite matrix satisfying the condition $V_{AB}=W_{AC}W^C_B$.  Operators $(p_{A}+iW_{AC}q^{C})$ and $(p_{B}-iW_{BD}q^{D})$ are  annihilation  and  creation operators respectively, similar to that of  harmonic oscillator problem and they obey similar commutation relation,
\be
[a_A,a_B^{\dagger}]=2W_{AB}.
\ee
If $\psi_{0}$ is the ground state for the harmonic oscillator system, then it follows the condition 
\be
(p_{A}-iW_{AC}q^{C})|\psi_{GS}>=0
\ee
and the solution is given by, \cite{Bombelli}
\ba
\psi_{GS}(\{q^C\})&=&<\{q^C\}|\psi_{GS}>\nn\\
&=&\left[\det\frac{W}{\pi}\right]^{1/4} \mathrm{exp} \left[-\frac{1}{2}W_{AB}\,q^{A}\,q^{B}\right].
\ea
The density matrix of the ground state is obtained by
\begin{align}
\rho\left(\{q^A\},\{q^{\prime B}\}\right)&=<\{q^A\}|0><0|\{q^{\prime B}\}>\nn\\
&=\left[\mathrm{det}\frac{W}{\pi}\right]^{1/2}\mathrm{exp}\left[-\frac{1}{2}W_{AB}\,(q^{A}\,q^{B}+q^{\prime A}\,q^{\prime B})\right]
\end{align}
We split ${q^{A}}$ into two subsystems, ${\{q^{a}\}}$ $(a=1,2,.......n_B)$ and $\{q^\alpha\}$ $(\alpha=n_B+1,n_B+2,.......N)$\footnote{The subsystem $\{q^a\}$ and subsystem  $\{q^{\alpha}\} $ regards as the inside and outside mode of the horizon}. The reduced density matrix of the subsystem ``1'' is obtained by tracing the degrees of freedom of the subsystem ``2'' \footnote{The subsystem ``1'' and subsystem ``2'' refers to the  subsystem with label ``$a$'' and  subsystem with label ``$\alpha$'' respectively. }, and is given by; 
\be
\rho\Big(\{q^{\prime a}\},\{q^{\prime b}\}\Big)=\int\prod_{\alpha}dq^{\alpha}\rho\,\Big(\{q^a,q^{\alpha}\},\{q^{\prime b},q^{\alpha}\}\Big)
\ee
 The matrix W naturally splits into four blocks as\cite{DS2,DS3,DS4},
\[ (W)_{AB} = \left(\begin{array}{ccc}
A_{ab} & B_{a\beta} \\
B^T_{\alpha b} & D_{\alpha \beta}\end{array} \right).\]
Now we find that reduced density matrix can be written as,
\begin{align}
\rho_{red}(\{q^{a}\},\{q^{\prime b}\})=&\left[\mathrm{det}\frac{M}{\pi}\right]^{1/2}\mathrm{exp}\left[-\frac{1}{2}M_{ab}(q^{a}q^{b} +q'^{a}q'^{b})\right]\nn\\ 
&~~~~~\mathrm{exp}\left[-\frac{1}{4}(N)_{ab}(q-q')^a (q-q')^b\right],
\label{rdm}
\end{align}
where 
\ba
M_{ab}=(A-BD^{-1}B^{T})_{ab} \qquad\text{and}\qquad N_{ab}=(B^TA^{-1}B)_{ab}.
\ea
The reduced density matrix of the system `1' is obtained by tracing the degrees of freedom of the system `2' and is same as above equation (\ref{rdm}).
The system can be diagonalized by the unitary matrix $U$ and the transformations
\be
q^{a}\rightarrow \tilde{q}^a=(UM^{1/2})^a_bq^{b}.
\ee
 Thus the density matrix reduces to \cite{Bombelli}, 
\begin{align}
\rho_{red}(\{q^{a}\},\{q^{b}\})&=\Pi_{n}\left[\pi^{-1/2}\exp\left(-\frac{1}{2}(q_n q^n+q^{\prime}_n q^{\prime n}\right.\right.\nn\\
&~~~~~~~~~~-\left.\left.\frac{1}{4}\lambda_i (q-q^{\prime})_n (q-q^{\prime})^n)\right)\right],
\end{align}
where $\lambda_i$  are the eigenvalues of the matrix $\Lambda^a_b=(M^{-1})^{ac}N_{cb}$. The entropy of the system can be calculated by the relation (25).

\section*{Acknowledgements}

I would like to thank Dr. Sanjay Siwach for useful discussions. The work of Dharm Veer Singh is supported by Rajiv Gandhi National Fellowship Scheme of University Grant Commission (Under the fellowship award no. F.14-2(SC)/2008 (SA-III)) and of Shobhit Sachan is supported by Council of Scientific and Industrial Research (CSIR) (Under the fellowship award no. 09/013(0239/2009-EMR-1)) of Government of India.

\end{document}